\documentclass[english,prd,superscriptaddress,nofootinbib,
               preprintnumbers,twocolumn,showpacs]{revtex4}

\usepackage[latin1]{inputenc}
\usepackage{amsmath}
\usepackage{amssymb}
\usepackage{graphicx}
\usepackage[caption=false]{subfig}
\usepackage{amsthm}
\usepackage{bm}
\usepackage{color}

\makeatletter
\@ifundefined{textcolor}{}
{
 \definecolor{BLACK}{gray}{0}
 \definecolor{WHITE}{gray}{1}
 \definecolor{RED}{rgb}{1,0,0}
 \definecolor{GREEN}{rgb}{0,1,0}
 \definecolor{BLUE}{rgb}{0,0,1}
 \definecolor{CYAN}{cmyk}{1,0,0,0}
 \definecolor{MAGENTA}{cmyk}{0,1,0,0}
 \definecolor{YELLOW}{cmyk}{0,0,1,0}
 }

\usepackage{amsfonts}\usepackage{dcolumn}

\def\be{\begin{equation}}
\def\ee{\end{equation}}
\def\ba{\begin{eqnarray}}
\def\ea{\end{eqnarray}}
\def\bs{\begin{subequations}}
\def\es{\end{subequations}}

\newcommand{\bfx}{}

\usepackage{babel}
\begin{document}

\title{An exact quantification of backreaction in relativistic cosmology}

\author{Timothy Clifton}

\affiliation{Department of Astrophysics, University of Oxford, UK.}

\affiliation{School of Physics and Astronomy,
 Queen Mary University of London, 
Mile End Road, London E1 4NS, UK.}

\author{Kjell Rosquist}

\affiliation{Department of Physics, Stockholm University, 106 91
  Stockholm, Sweden.}
  
\affiliation{ICRANet, Piazza della Repubblica, 10  I--65122 Pescara, Italy}

\author{Reza Tavakol}

\affiliation{School of Physics and Astronomy,
 Queen Mary University of London, 
Mile End Road, London E1 4NS, UK.}

\begin{abstract}

An important open question in cosmology is the degree to which the
Friedmann-Lema\^{i}tre-Robertson-Walker (FLRW) solutions of Einstein's
equations are able to model the large-scale behaviour of the locally inhomogeneous
observable universe.  We investigate this problem by considering
a range of {\it exact} n-body solutions of Einstein's constraint equations. These 
solutions contain discrete masses, and so allow arbitrarily large density
contrasts to be modelled. We restrict our study to regularly arranged
distributions of masses in topological 3-spheres.  This has the
benefit of allowing straightforward comparisons to be made with FLRW
solutions, as both spacetimes admit a discrete group of symmetries.
It also provides a time-symmetric hypersurface at the moment of
maximum expansion that allows the constraint equations to be solved
exactly.  We find that when all the mass in the universe is condensed
into a small number of objects ($\lesssim 10$) then the amount of
backreaction in dust models can be large, with $O(1)$ deviations from the
predictions of the corresponding FLRW solutions. When the number of
masses is large ($\gtrsim 100$), however,
then our measures of backreaction become small ($ \lesssim 1\%$).
This result does not rely on any averaging procedures, which are
notoriously hard to define uniquely in general relativity, and so provides (to
the best of our knowledge) the first exact and unambiguous
demonstration of backreaction in general relativistic cosmological modelling.
Discrete models such as these can therefore be used as laboratories to test ideas
about backreaction that could be applied in more complicated and
realistic settings.

\end{abstract}
\maketitle
\section{Introduction}

The visible matter in the universe contains structure on a wide range
of scales, from stars and planets ($\sim 10^{10}$m) to
voids and super-clusters ($\sim 10^{24}$m), ({\bfx see e.g.\  \cite{Ellis-1984}}).  In standard general
relativistic cosmological modelling the problem of dealing with this
complexity is often circumvented by assuming that the stress-energy
tensor of an inhomogeneous universe can be approximated by an `averaged' stress-energy
tensor, usually assumed to be representable by a simple perfect fluid,
while leaving the left hand side of Einstein's equations unchanged.  The
solution to these equations is then taken to represent the geometry of
spacetime on the scale over which the averaging was performed. If
this scale is large enough to result in a homogeneous and isotropic
spacetime, then we will refer to this geometry as the ``background''.

The great benefit of this approach is the enormous simplification
it entails, by  allowing cosmology to be done on any scale, without having to worry about the enormously complicated task of taking into account the wide range of structures that exist in the universe.  The drawback is that it is not always clear to what extent the expansion of space in the background geometry is representative of the large-scale expansion of space in the actual universe.  For a given model, we will refer to the difference between these two things as ``backreaction'', which can be considered as the influence of inhomogeneities at different scales on large-scale cosmological dynamics.  
As large-scale expansion can be related to observables such as luminosity distance and redshift \cite{rasa,rasb,clif}, it is therefore the case that an understanding of backreaction could prove essential in achieving a fundamental concordance between theory and observations \cite{ras1, inhomorev}.

This task becomes increasingly important given that observational
cosmology is now starting to be able to make high-precision
measurements of a variety of different astrophysical probes, including
the Cosmic Microwave Background (CMB) \cite{wmap,spt}, Type Ia
supernovae \cite{snls,sdss1}, and Baryon Acoustic Oscillations (BAO)
\cite{sdss2,2df}.  When interpreted within the background of the standard
cosmological model, these very different probes all suggest that the
universe  is at present undergoing a phase of late-time accelerated
expansion.  In a genuinely homogeneous and isotropic general
relativistic universe filled with a perfect fluid, such behaviour is
only possible if $p<-\rho/3$, with data currently being compatible with 
{\bfx the presence of a cosmological constant with} $p= -\rho$. 
The presence of such a fluid is problematic for a variety of
reasons, requiring a fine-tuning of perhaps as much as $1$ part in
$10^{120}$.  This immediately raises the question of whether the model
we are using to interpret the data is adequate, a question which needs
to be quantified in any case if we are to start performing ``precision
cosmology'' \cite{precision}.

One particular {\bfx aspect of this} complicated problem is the degree to
which dust (a fluid without any self-interactions) is appropriate for describing the matter
content of the universe \cite{wilt}.  This approximation treats gravitationally bound,
extended objects (such as galaxies, and clusters of galaxies) as
a medium that is parametrised by only a handful of continuous,
and usually slowly varying, quantities. Within this framework, the spacetime
through which photons travel has a fundamentally different type of
curvature to that of the real universe \cite{bert}. It also ignores the
gravitational consequences of the binding energies between and within
astrophysical systems \cite{kjell}.  A dust description should
therefore be considered as an approximation only, and
must itself be investigated in order to ascertain the extent to which it is
justifiable.
 
We study the problems outlined above by considering cosmological solutions that
are explicitly composed of regularly arranged discrete masses.  These
solutions consist of vacuum everywhere exterior to the sources
under consideration, and are constructed on a manifold that has
spacelike hypersurfaces that are topological 3-spheres.  As a result, 
they admit a group of discrete symmetries corresponding to
rotations of the 3-sphere that leave the positions of the masses
invariant.  These models are among the simplest possible
inhomogeneous spacetimes that could be considered to exhibit
large-scale homogeneity and isotropy:  They introduce only one extra scale into
the problem (the inter-particle separation), they are vacuum
solutions of the field equations (up to the singularities),
and they contain within them a time-symmetric spatial hypersurface
(which greatly simplifies the constraint equations).  This makes them
ideal candidates for studying backreaction. 

Cosmological solutions of this type have been considered before, starting
with the seminal work of Lindquist and Wheeler \cite{LW}.  The
approach these authors took was to construct a gravitational
analogue of the highly successful Wigner-Seitz construction.  The basic idea
was to build lattices from regular
cells, and attach the cell boundaries tangentially to a background
3-sphere.  Dynamics were then imposed by assuming that observers that
are equidistant between neighbouring sources should be falling
freely with respect to those sources. This led to a dynamical model
of the universe in which the masses are regularly arranged, and that is taken to be an
approximate solution to Einstein's equations. The Lindquist-Wheeler approach has 
recently been extended to include models with spatially flat and open topologies
\cite{Clifton:2009jw}, as well as to models with a cosmological constant 
\cite{next}.  The optical properties of spatially flat models of this
type have also been considered in \cite{Clifton:2009jw,next,next2}, 
as have those of negatively curved models in \cite{Kane}.  Other
attempts to construct cosmological models for similar configurations
of discrete masses have been made using perturbative expansions
\cite{noaveraging}, and Regge calculus \cite{Collins-Williams-73}.

While providing insights into the problem, all of the models mentioned
above involve the use of approximate solutions.  Our aim here is to
take an alternative approach to addressing this problem that is exact.
This approach is based on work initiated by Misner {\bfx about five decades}
ago \cite{Misner-63}, and that has subsequently been used
extensively in the study of black hole physics
\cite{Brill-Lindquist-63, Gibbons, Cadez}.  The underlying idea here
is again drawn from analogy 
with the study of electromagnetism, where intuitions 
about more general situations are gained by first studying the simple case of static configurations of charge.  In
gravitational settings, however, it is very difficult to arrange for
configurations of isolated masses to be static for a finite interval of
time\footnote{Counter-examples include spacetimes in which the
cosmological constant is non-zero, and the periodic solution found
by Korotkin and Nicolai \cite{KN}.}.  In contrast, configurations
that are {\it instantaneously} static occur
frequently. The study of such configurations, that occur in many systems of
physical interest, is referred to as geometrostatics\footnote{Examples
  include dynamical systems that possess a recollapsing phase (as
  happens when a particle is thrown vertically upwards from the surface of a
massive body, with less than the escape velocity).}.

In the case of cosmology, which is our main interest here,
an important example is provided by the instantaneously static (and
therefore time-symmetric) hypersurfaces that occur at the
moment of maximum expansion in dust-dominated FLRW solutions with
positive spatial curvature.
Our goal is to construct a set of instantaneously static configurations that are exact, 
and that contain discrete matter sources, rather than continuous fluids. 
We do this by constructing regular tessellations of a 3-sphere by using
different sets of identical polyhedra, 
each with an identical mass at their centre.  We show that the method
of geometrostatics can be extended to solve the constraint equations
{\em exactly} in these cases.
Our aim is then to use these solutions to study backreaction in dust filled
cosmological models.  We do this by comparing 
the {\bfx time-symmetric spatial hypersurfaces of the exact discrete solutions
to the corresponding} FLRW solutions with the same {\bfx proper
mass\footnote{{\bfx As defined in \cite{Wald:1984} and in Section \ref{msm}.}}},
and considering, 
in particular, the extent to which the FLRW solutions emerge as the number of sources is increased. 
This provides a {\it precise} indication of the degree to which FLRW solutions can be used to 
represent initial configurations for spacetimes that contain discrete
objects.  A fuller investigation of backreaction along these
lines would also include considering the dynamical evolution of
models with discrete sources.  We leave this for a future publication.

The plan of the paper is as follows. In Section \ref{2} we recap on
perfect fluid FLRW cosmology.  In Section \ref{equations} we provide
the constraint equations on time-symmetric hypersurfaces for discrete
masses on a 3-sphere.  We discuss the Schwarzschild solution,
and then outline how to find the solution for arbitrarily many discrete
masses in a closed space.  In Section \ref{Sec1} we construct all of
the possible solutions that consist of discrete sources regularly
arranged on a closed lattice.  This allows us to explain why there
are no 2-mass solutions with spherical topology to Einstein's equations
that admit a time-symmetric hypersurface.  In Section \ref{Sec4} we
provide evidence that shows that the sources in our discrete models
are always separated by distances that are larger than their horizon sizes at the
maximum of expansion.  In Section \ref{Sec2} we investigate backreaction
in dust-dominated cosmological models by comparing
the scale of our lattices to dust-filled FLRW solutions of
Einstein's equations with the same total proper mass.  We find that $O(1)$
deviations from the results of Friedmann cosmology can
occur when the number of masses is small ($\lesssim 10$), but
that the scale of the solutions converge to the Friedmann values when there are very many
masses ($\gtrsim 100$).  In Section \ref{Sec3} we then consider what
happens when we include inter-particle interaction energies in our
definition of mass.  We find that the consequences of changing the
number of masses in the lattice, while keeping the total energy of the
system the same, can considerably change the scale of the hypersurface
of maximum expansion.  This shows that although backreaction in
dust models we consider is generically quite small, the consequences
of ignoring interaction energies when using the dust approximation can
hide potentially interesting effects.  In Section \ref{conclusion} we present
our conclusions.

\section{Perfect fluid cosmology}
\label{2}

The spatially homogeneous and isotropic perfect fluid FLRW solutions
have a geometry given by
\be
ds^2= -f(t) dt^2 +a^2(t) \left( d \chi^2 + h^2(\chi) d \Omega^2
\right), 
\ee
where $f(t)$ is a free function, $d\Omega^2=d\theta^2
+\sin^2(\theta)d\phi^2$, and $h(\chi) = \sin \chi$, $\chi$ or $\sinh
\chi$ for solutions with spatial
curvature $k=+1$, $0$ or $-1$, respectively.  The scale factor $a(t)$
satisfies the constraint equation
\be
\frac{1}{f} \frac{\dot{a}^2}{a^2} = \frac{8 \pi}{3} \rho -
\frac{k}{a^2},
\ee
where the dot represents a derivative with respect to $t$, and
$\rho=\rho(t)$ is the energy density of the continuous perfect fluid,
which obeys the conservation equation
\be
\dot{\rho} +3 \frac{\dot{a}}{a} \left(\rho+p\right) =0,
\ee
where $p=p(t)$ is the pressure.  This completely specifies the
solution, up to constants of integration.

For the present study we will be primarily interested in models with
positive spatial curvature ($k=+1$), and with a pressureless dust
source ($p=0$).  In this case, if we take $f=a(t)$, then the solution
can be written as
\be
a(t) = \frac{8 \pi}{3} \rho_0 -\frac{1}{4} (t-t_0)^2,
\ee
where $t_0$ and $\rho_0$ are constants, and where
$\rho(t)=\rho_0/a^3(t)$. The maximum of expansion can then be seen to
occur at $t=t_0$, and the geometry of the hypersurface $t=t_0$ can be
seen to be given by
\be
\label{FLRWmax}
dl^2 = \frac{3}{8 \pi \rho(t_0)} \left( d \chi^2 +
\sin^2\!\chi \, d \Omega^2 \right),
\ee
which is rigidly specified once $\rho(t_0)$, the energy density at
maximum of expansion, is known.

We note that the only dust-filled FLRW solutions that admit a time-symmetric 
hypersurface are spatially closed, with a spherical topology. 
Interestingly, the boundary conditions for discrete models with a momentarily 
static distribution of sources also seem to be incompatible with open
topologies (the reason for this is that a $1/r$ source cannot live
alone on $T^3$, and a regular lattice of identical sources on $E^3$
is also not possible \cite{KN}). This shows a certain qualitative
concordance between discrete and fluid solutions in the interplay between
boundary conditions and topology.

\section{Constraint equations for discrete models}
\label{equations}
Our aim here is to obtain exact vacuum solutions of Einstein's equations
corresponding to regular lattices of sources that are instantaneously
at rest on a topological 3-sphere.

In such a setting the relevant equations to solve are the
Gauss-Codazzi equations:
\begin{eqnarray}
\label{GC1}
\mathcal{R}+K^2-K_{ij} K^{i j} &=&0\\[3pt]
\label{GC2}
\bigl( K_{i}^{\phantom{i}
  j}-\delta_{i}^{\phantom{i} j} K\bigr)_{\vert j}
&=&0,
\end{eqnarray}
where $\mathcal{R}$ is the Ricci curvature of the 3-space, $K_{i
  j}$ is the extrinsic curvature of the 3-space in the
4-dimensional spacetime, and $K=
K_{i}^{\phantom{i}i}$.  The indices $i$, $j$
refer to coordinates in the 3-space, and the vertical line denotes
covariant derivative in that space.

It is well known that if we choose a time coordinate that is
specified by the normal derivative to this initial hypersurface
(such that $g_{t\mu} = -\delta_{\mu}^{\phantom{\mu} t}$), then the
extrinsic curvature can be written as $K_{i j} = -\frac{1}{2}
g_{i j,t}$. Instantaneously static hypersurfaces therefore
have $K_{i j}=0$, and the Gauss-Codazzi equations reduce simply to
\be
\label{GC3}
\mathcal{R}=0.
\ee
A key point here is that for any 3-dimensional geometry that satisfies 
the initial constraint (\ref{GC3}) there is
a unique 4-dimensional spacetime that satisfies the full
Einstein equations.  This is the method of geometrostatics,
presented by Misner in \cite{Misner-63}, and studied for the case of
asymptotically flat space in \cite{Brill-Lindquist-63, Gibbons, Cadez}.

Here we are interested in solving Eq.~(\ref{GC3}) in closed spaces.
In this case we can make the following ansatz for the metric of the spatial 3-section
\be
\label{Hh1}
dl^2 = \psi^4 \hat{h}_{i j} dx^{i} dx^{j},
\ee
where $\psi=\psi(x^{i})$, and where $\hat{h}_{i
  j}$ is the metric of a 3-sphere with constant curvature
$\hat{\mathcal{R}}$.  The Gauss-Codazzi equations are
then satisfied if $\psi$ obeys the Helmholtz equation:
\be
\label{Hh}
\hat{\nabla}^2 \psi = \frac{1}{8} \hat{\mathcal{R}} \psi,
\ee
where $\hat{\nabla}^2$ is the Laplacian corresponding to
$\hat{h}_{i j}$.  

\subsection{A Single Schwarzschild Mass}

The Schwarzschild solution has been well studied using
asymptotically flat solutions in geometrostatics, including in the
original work of Misner \cite{Misner-63}.  It is important to note
that the Schwarzschild solution is still a solution in the present case,
as the 3-sphere is conformally flat.  In this case, 
however, there will be a different radial variable and a different functional form
for the conformal factor $\psi$.  Given the linearity of the Helmholtz equation,
\eqref{Hh}, it can be seen that multi-source solutions
can be constructed by linear superposition of the solutions for
single sources.

In asymptotically flat, isotropic coordinates, a static spacelike
slice of the Schwarzschild solution can be written as
\be
\label{schw1}
dl^2 = \left(1 +\frac{m}{2 \tilde r} \right)^4 (d\tilde r^2 + \tilde r^2
d\Omega^2),
\ee
where $m$ is a constant.  Performing the coordinate transformation
$\tilde r =K \tan \frac{\chi}{2}$, where $K$ is a constant, it can be
seen that Eq.~(\ref{schw1}) becomes
\be
\label{Schw-spherical}
dl^2= 
\frac{K^2}4 \left[ \frac1{\cos\frac\chi2} +
                    \frac{m}{2K\sin\frac\chi2} \right]^4
                (d\chi^2 + \sin^2\!\chi d\Omega^2).
\ee
This metric is clearly of the same form as Eq. (\ref{Hh1}), and solves
Eq.~(\ref{Hh}), with $\hat{\mathcal{R}}=6$ (for a unit 3-sphere) and
\be
\psi = \sqrt{\frac{K}{2}} \left( \frac1{\cos\frac\chi2} +
                    \frac{m}{2K\sin\frac\chi2} \right).
\ee
Actually, both functions $A(\chi):= (\sin\frac\chi2)^{-1}$ 
and $B(\chi):= (\cos\frac\chi2)^{-1}$  satisfy Eq.~\eqref{GC3} with
$\hat{\mathcal{R}}=6$. 

This means that although the term in Eq.~(\ref{Schw-spherical}) that corresponds directly to
the $2 m/r$ source term from Eq.~(\ref{schw1}) can be seen to be the one
that is proportional to $A(\chi)$, the term that is proportional to $B(\chi)$ can
also be treated as a source, although it is located at the antipode.
This becomes obvious if we place the origin of the spherical
coordinates at the antipode by the transformation $\chi \rightarrow
\pi-\chi$, as in that case the roles of the two terms in
Eq.~(\ref{Schw-spherical}) are interchanged (i.e.\ $A(\chi) \rightarrow
B(\chi)$ and $B(\chi) \rightarrow A(\chi)$).
Thus it appears that placing a Schwarzschild source on the 3-sphere induces 
a mirror source at the antipode. Furthermore, these two source terms are 
joined at their horizons, as is clear if we set the gauge parameter to
$K=m/2$ (in which case the horizon at $\tilde r =m/2$ appears at
$\chi=\pi/2$ in the hyperspherical coordinates).

One can then consider that as $\chi
\rightarrow \pi$ one approaches either the asymptotic region where $\tilde{r}\rightarrow
\infty$, {\it or} that one approaches another Schwarzschild mass.
These two situations are geometrically identical, and there is
therefore no difference between the one mass solutions and the two
mass solution (as long as the point at $\chi = \pi$ can be added to
the manifold).

\subsection{Many Schwarzschild Masses}
\label{msm}

From Eq.~\eqref{Schw-spherical} we can now infer that there exist
multi-source solutions on the 3-sphere that take the form
\begin{equation}\label{sphere_general}
 dl^2 = \psi^4 (d\chi^2+\sin^2\!\chi\,d\Omega^2), \\[5pt]
\end{equation}
where $\psi = \psi(\chi, \theta,\phi)$ is given by
\begin{equation}
 \psi(\chi, \theta,\phi)
       = \sum_{i=1}^N \frac{\sqrt{\tilde m_i}}{2f_i(\chi, \theta, \phi)} \ ,
\end{equation}
where the mass parameters $\tilde m_i$ are a set of constants and
$f_i(\chi, \theta, \phi)=\sin({\chi_i}/2)$.  Here the $\chi_i$ refer
to new coordinates $(\chi_i, \theta_i, \phi_i)$ which are obtained by
rotating the coordinates $(\chi,\theta,\phi)$ in 
Eq.~(\ref{sphere_general}) so that the $i$'th source position appears at
${\chi_i}=0$.

It follows from the form of Eq. \eqref{sphere_general} that one cannot
adjust the $\tilde m_i$ and the size of the 3-sphere independently.  For
example, scaling the size of the 3-sphere by a constant $\alpha^2$
automatically results in scaling the $\tilde m_i$ by a factor $\alpha$. This
demonstrates that there exists a certain rigidity in the configuration,
once the $\tilde m_i$ are specified. Specifying the value of the $\tilde{m}_i$ is
therefore crucial for determining any measure of distance in our
hypersurface.  Once they are specified, however, then all measures
of distance can be calculated uniquely.

When considering the original single source Schwarzschild solution,
the parameters $\tilde m_i$ ($\tilde m_1$ and $\tilde m_2$ in that
case) are equal to the standard Schwarzschild mass, $\tilde m_1=
\tilde m_2=m$ (using the gauge $K=m/2$). However, when multiple
sources are present, the interpretation of the $\tilde m_i$ parameters
will instead be that of an  \emph{effective mass} which includes the
binding energies with respect to all the other objects in the universe
(cf.~\cite{Brill-Lindquist-63}). 
However, their actual, locally measured, \emph{proper
  mass}\footnote{This term is used in analogy with the proper mass
  defined by Wald \cite[p.126]{Wald:1984}. In the context of
  asymptotically flat spacetimes, Brill and Lindquist
  \cite{Brill-Lindquist-63} use the term \emph{bare mass}, as it is used
  in particle physics. We prefer not to use this term here since the
  situations in cosmology and particle physics are opposites with
  respect to how masses are measured. In particle physics, the
  observer is outside the system and cannot measure the bare mass
  directly. In cosmology, on the other hand, the observer is inside
  the system and therefore measures the bare mass (in particle physics
  terminology).} must be determined by the requirement that the
geometry reduces to that given by  Eq.~(\ref{Schw-spherical}) in the
limit $\chi \rightarrow 0$, in analogy with the asymptotically flat
situation analysed in Ref.\ \cite{Brill-Lindquist-63}. It can be
shown that the proper mass defined in this way is equivalent to the
mass in the Schwarzschild region at the other end of the
Einstein-Rosen bridge of the black hole. 

To summarise, these two mass definitions are:
\begin{enumerate}
\item[1] {\it The Effective Mass}, $(m_i)_\text{eff}\,$, is defined to
  be equal to the mass parameter of the corresponding source,
  $(m_i)_\text{eff} := \tilde m_i$. 
\item[2] {\it The Proper Mass}, $m_i$, is defined in such a way
  that the geometry of the space in the vicinity of each object is
  given by Eq.~(\ref{Schw-spherical}), in the limit $\chi \rightarrow 0$.
\end{enumerate}
Strictly speaking, one requires a region to exist that is infinitely far away from all of
the masses in order to have an
operational definition of the gravitational effect of the
inter-particle potentials that are included in the effective mass.  This is clearly impossible in the case
of a closed space, but this does not stop us from drawing an analogy with
cases in which asymptotically distant regions do exist.

\section{Lattice solutions}
\label{Sec1}

Using the above ingredients we can now construct 
a set of models with different numbers of regularly arranged discrete
masses on a 3-sphere.  To do this we proceed in analogy 
with the approach taken by Lindquist and Wheeler \cite{LW}, so that
our models are constructed by considering all of the possible regular tessellations
of the 3-sphere.  We then place identical masses at the
centre of each cell, with the result that the distribution of
masses is such that the distance between any mass and its nearest
neighbours is the same for each of them.

There are six possible regular tessellations of the 3-sphere,
those with 5, 8, 16, 24, 120 and 600 polyhedra \cite{tiling}.
In addition, one can also cover a hypersphere with 2 balls
\cite{Uzan-etal}.  These possibilities are displayed in Table
\ref{table1}.  Using these tessellations, we will now proceed to 
construct a sequence of exact discrete regular
lattice models of  the universe, with increasing numbers of sources, on
a time-symmetric 3-sphere.
Our discrete models will consist of $n$ equal masses (where
$n=2,5,8,16,24,120,600$).  By positioning these masses at the centre of each of the
cells, they are at a maximum possible distance away from
each of their neighbours.  

\begin{table}[htb]
\begin{center}
\begin{tabular}{|c|c|c|c|}
\hline
$\begin{array}{c} \bf{Lattice}\\
  \bf{Structure} \end{array}$ & 
 $\begin{array}{c} \textbf{Cell}\\
  \textbf{Shape} \end{array}$
& $\begin{array}{c} \textbf{Number of}\\
  \textbf{Cells} \end{array}$ \\
\hline
- & Ball & 2 \\
\{333\}  & Tetrahedron & 5 \\
\{433\}  & Cube & 8 \\
\{334\}  & Tetrahedron & 16 \\
\{343\}  & Octahedron & 24 \\
\{533\}  & Dodecahedron & 120 \\
\{335\}  & Tetrahedron & 600 \\
\hline
\end{tabular}
\end{center}
\caption{
\textit{All possible regular tessellations of the 3-sphere.
The `Lattice Structure' is given by the Schl\"{a}fli symbols $\{pqr\}$, where
$p$ is the number of edges to a face, $q$ is the number of faces 
that meet at a vertex, and $r$ are the number of cells that meet at 
an edge \cite{tiling}.}}
\label{table1}
\end{table}

\subsection{The 2-Cell model}

The 2-cell model, consisting of two balls with their boundaries 
identified, is the exceptional case of the structures listed in Table
\ref{table1}.  Its cells are not regular polyhedra (and hence it has
no Schl\"{a}fli symbol), and it was not considered in either \cite{LW} or
\cite{tiling}.  Nevertheless, a lattice with 2 cells seems like a
perfectly legitimate object to consider\footnote{This object does, 
in fact, have a considerable advantage over the lattices in 
Table \ref{table1}, as the geometry around each mass must be 
spherically symmetric, and so is considerably easier to solve for.
This means that the approximations used by Lindquist and Wheeler 
in \cite{LW} are not required in this case at all.} and has indeed 
been studied recently in \cite{Uzan-etal}. The authors of this
paper found that the geometry inside each cell could only be matched
at the junction between cells if that junction was a horizon.  They
also found that no solutions exist unless the cosmological constant
$\Lambda \neq 0$.

Let us now consider this 2-cell model using the initial value formalism
discussed above.  The source functions in this case are given by $f_1=
\sin(\chi/2)$ and $f_2= \cos(\chi/2)$, with $\tilde m_1=\tilde m_2=m$.
Substituting into  Eq.~(\ref{sphere_general}) then gives
\begin{equation}
   \psi = \frac{\sqrt{ m}}{2\cos\frac\chi2}
        + \frac{\sqrt{m}}{2\sin\frac\chi2} \ ,
\end{equation}
which is clearly just the Schwarzschild solution given in
Eq.~(\ref{Schw-spherical}), with $K=m/2$.  As explained above, this model can be
considered to consist of two sources centred at opposite poles and
matched (analytically continued in fact) across their horizons. Within
this interpretation the exterior parts of the sources are missing, and
there are consequently no static regions outside of the black
holes. Because of the absence of an exterior region between the
sources, the 2-cell model is unsuitable for cosmology. This result
provides an alternate illustration of the findings in
Ref.~\cite{Uzan-etal}.
 
\subsection{The 5-Cell model} 

Let us now consider the 5-cell model (or {\it hyperpyramid}), that consists
of 5 tetrahedra.  This structure, which is in fact a 4-simplex, is the
tessellation in Table \ref{table1} with the fewest number of cells that are
regular polyhedra.  To obtain the coordinates of the masses
at the centres of the tetrahedra one can consider a hyperpyramid in
the embedding space $E^4$, and place a unit 3-sphere inside the
hyperpyramid such that the two structures touch at the centre of each of the 5
tetrahedral cells. Alternatively, one could consider the positions of
the vertices of the dual lattice\footnote{Dual lattices have the
  vertices and centres of each cell transposed with each other.}, which in this case is another
hyperpyramid. The coordinates of the 5 masses that result are
given in the Table \ref{table3}, below.

To define the model we use spherical polar coordinates which are
related to Cartesian coordinates of $E^4$ by 
\begin{equation}
 \begin{split}
 w &= \cos\chi\\
 x &= \sin\chi \cos\theta \\
 y &= \sin\chi \sin\theta \cos\phi \\
 z &= \sin\chi \sin\theta \sin\phi \ .
 \end{split}
\end{equation}

The source functions, $f_i$, from Eq.~(\ref{sphere_general}), are then
\be
\label{fi2}
f_i = \sin \left[ \textstyle\frac{1}{2} \cos^{-1} (h_i)\right],
\ee
where the functions, $h_i$, defined in Eq.~(\ref{fi2}) are given by 
\begin{eqnarray*}
h_1&=&\cos\chi\\
h_2&=&\frac{\sqrt{15}}{4}
  \cos \theta \sin \chi -\frac{\cos \chi}{4}\\
 h_3&=&\sqrt{\frac{5}{6}}
  \sin \chi \sin \theta \cos \phi \\&&-\sqrt{\frac{5}{48}} \sin
  \chi \cos \theta -\frac{\cos \chi}{4}\\
h_4&=&\sqrt{\frac{5}{6}}
  \sin \chi \sin \theta \sin \left(\phi-\frac{\pi}{6} \right)
  \\&&-\sqrt{\frac{5}{48}} \sin  \chi \cos \theta -\frac{\cos \chi}{4}\\
h_5&=&-\sqrt{\frac{5}{6}}
  \sin \chi \sin \theta \sin \left(\phi+\frac{\pi}{6} \right)
  \\&&-\sqrt{\frac{5}{48}} \sin \chi \cos \theta -\frac{\cos \chi}{4} .
\end{eqnarray*}
The geometry of this model can be visualised by considering a slice 
through it. To this end, consider the surface 
$\chi=\chi_0$ for $\chi_0 = \cos^{-1}(-1/4) \approx 1.82$. Its metric is given by
\begin{equation}\label{subspace}
   dl^2 = \frac{15}{16} \,\psi^4(\chi_0, \theta, \phi) \, d\Omega^2
\end{equation}
To get a rough idea of the shape of this hypersurface
we can think of $\theta$ and
$\phi$ as polar angles in $E^3$ and plot the surface $\psi(\chi_0,
\theta, \phi) = \rho$  where $\rho^2 = x^2 + y^2 + z^2$. This surface
is displayed in Fig.~\ref{5tetrafig} and goes through 4 of
the 5 masses. It should be noted that the surface does not represent
an isometric embedding of the geometry in
Eq.~(\ref{subspace}). However, in regions where the derivatives of
$\psi$ are small, it does give an approximate representation of that
geometry. This approximation is therefore best in regions which are
far from the sources.

\begin{table}[htb]
\begin{center}
\begin{tabular}{|c|l|l|}
\hline
\bf{Point}  & \bf{($w$, $x$, $y$, $z$)} & \bf{($\chi$,
  $\theta$, $\phi$)}\\
\hline
$(i)$ & $(1,0,0,0)$  &  $\left(0, \frac{\pi}{2}, \frac{\pi}{2}
\right)$ \\
$(ii)$ & $\left(-\frac{1}{4},\frac{\sqrt{15}}{4},0,0\right)$  &
$\left(\cos^{-1}(-\frac14), 0, \frac{\pi}{2}
\right)$ \\
$(iii)$ & $\left(-\frac{1}{4},-\sqrt{\frac{5}{48}},
\sqrt{\frac{5}{6}},0\right)$  &  $\left(\cos^{-1}(-\frac14),
\cos^{-1}(-\frac13), 0
\right)$ \\
$(iv)$ & $\left(-\frac{1}{4},-
\sqrt{\frac{5}{48}},-\sqrt{\frac{5}{24}},
\sqrt{\frac{5}{8}} \right)$  
& $\left(\cos^{-1}(-\frac14),\cos^{-1}(-\frac13) , \frac{2 \pi}{3}\right)$ \\
$(v)$ & $\left(-\frac{1}{4},-
\sqrt{\frac{5}{48}},-\sqrt{\frac{5}{24}},-
\sqrt{\frac{5}{8}} \right)$  
& $\left(\cos^{-1}(-\frac14),\cos^{-1}(-\frac13) , \frac{4 \pi}{3} \right)$ \\
\hline
\end{tabular}
\end{center}
\caption{{\protect{\textit{Coordinates  {\rm ($w,x,y,z$)} of the 5 
masses in the embedding space $E^4$, as well as {\rm 
($\chi,\theta,\phi$)} on the background 3-sphere. In this table, and
throughout, $\cos^{-1}$ refers to the inverse cosine, and not its reciprocal.}}}}
\label{table3}
\end{table}

\subsection{The 8-Cell model}

We now proceed in a similar manner to find the geometry of the 8-cell
(or {\it tesseract}) model. In this case the primitive 
cell of our lattice is a cube. To find the position of the centre of each 
cell we can again embed  the structure in $E^4$, together with a unit 3-sphere.
Alternatively, the positions of the masses can be found using the dual
lattice, which in this case is the 16-cell. 
The positions of the masses are then given as in Table \ref{table2}.

The $f_i$ from Eq.~(\ref{sphere_general}) are then
found to be 
\begin{eqnarray*}
f_1 &=& \sin \left[\frac{\chi}{2}\right]\\
f_2 &=& \cos \left[\frac{\chi}{2}\right]\\
f_3 &=& \sin \left[ \frac{1}{2} \cos^{-1} \left( \cos \theta
  \sin \chi \right) \right]\\
f_4 &=& \cos \left[ \frac{1}{2} \cos^{-1} \left( \cos \theta
  \sin \chi \right) \right]\\
f_5 &=& \sin \left[ \frac{1}{2} \cos^{-1} \left( \cos \phi \sin \theta 
  \sin \chi \right) \right]\\
f_6 &=& \cos \left[ \frac{1}{2} \cos^{-1} \left( \cos \phi \sin \theta
  \sin \chi \right) \right]\\
f_7 &=& \sin \left[ \frac{1}{2} \cos^{-1} \left( \sin \phi \sin \theta
  \sin \chi \right) \right]\\
f_8 &=& \cos \left[ \frac{1}{2} \cos^{-1} \left( \sin \phi \sin \theta 
  \sin \chi \right) \right].
\end{eqnarray*}
The first two of these functions are identical to those corresponding 
to the 2-cell model, considered above.

We can again visualise the geometry of this model by considering a
slice though it.  In Fig.~\ref{8cubefig} we show the surface with
$\chi=\pi/2$, which passes through 6 of the 8 masses. The radial
position of the surface at any given ($\theta$, $\phi$) is
proportional to $\psi$.

\begin{table}[htb]
\begin{center}
\begin{tabular}{|c|l|l|}
\hline
\bf{Point}  & \bf{($w$, $x$, $y$, $z$)} & \bf{($\chi$,
  $\theta$, $\phi$)}\\ 
\hline
$(i)$ & $(1,0,0,0)$  &  $\left(0, \frac{\pi}{2}, \frac{\pi}{2}
\right)$ \\
$(ii)$ & $(-1,0,0,0)$  &  $\left(\pi, \frac{\pi}{2}, \frac{\pi}{2}
\right)$ \\
$(iii)$ & $(0,1,0,0)$  &  $\left( \frac{\pi}{2}, 0, \frac{\pi}{2}
\right)$ \\
$(iv)$ & $(0,-1,0,0)$  &  $\left( \frac{\pi}{2}, \pi, \frac{\pi}{2}
\right)$ \\
$(v)$ & $(0,0,1,0)$  &  $\left( \frac{\pi}{2}, \frac{\pi}{2}, 0
\right)$ \\
$(vi)$ & $(0,0,-1,0)$  &  $\left( \frac{\pi}{2}, \frac{\pi}{2}, \pi
\right)$ \\
$(vii)$ & $(0,0,0,1)$  &  $\left( \frac{\pi}{2}, \frac{\pi}{2}, \frac{\pi}{2}
\right)$ \\
$(viii)$ & $(0,0,0,-1)$  &  $\left( \frac{\pi}{2}, \frac{\pi}{2},
\frac{3 \pi}{2} \right)$ \\
\hline
\end{tabular}
\end{center}
\caption{{\protect{\textit{Coordinates  {\rm ($w,x,y,z$)} of the 8
masses in the embedding space $E^4$, as well as {\rm
($\chi,\theta,\phi$)} in the lattice.}}}}
\label{table2}
\end{table}

\subsection{Models with 16-600 Equally Spaced Masses}

\begin{figure*}
\centering
  \subfloat[A slice through the 5-cell solution]{\label{5tetrafig}
    \includegraphics[height=2.5in]{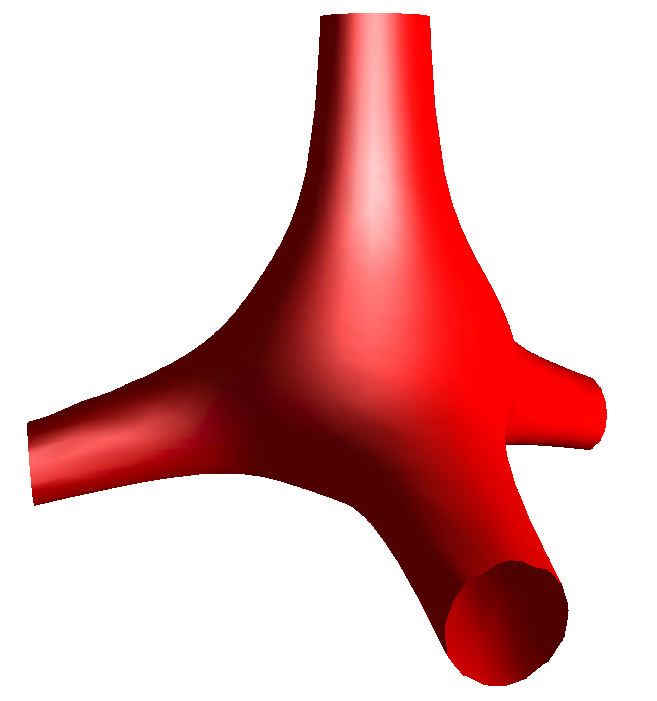}}\qquad\qquad
  \subfloat[A slice through the 8-cell solution]{\label{8cubefig}
    \includegraphics[height=2.5in]{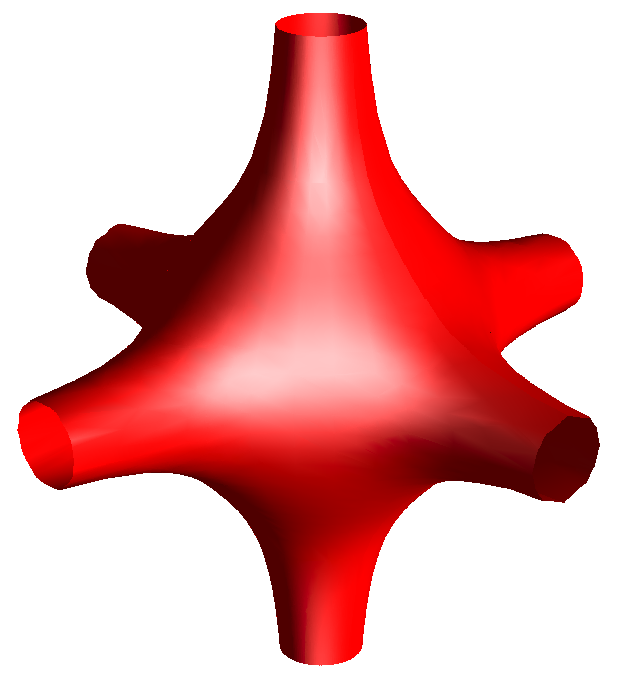}}\\
  \subfloat[A slice through the 16-cell solution]{\label{16cell}
    \includegraphics[height=2.5in]{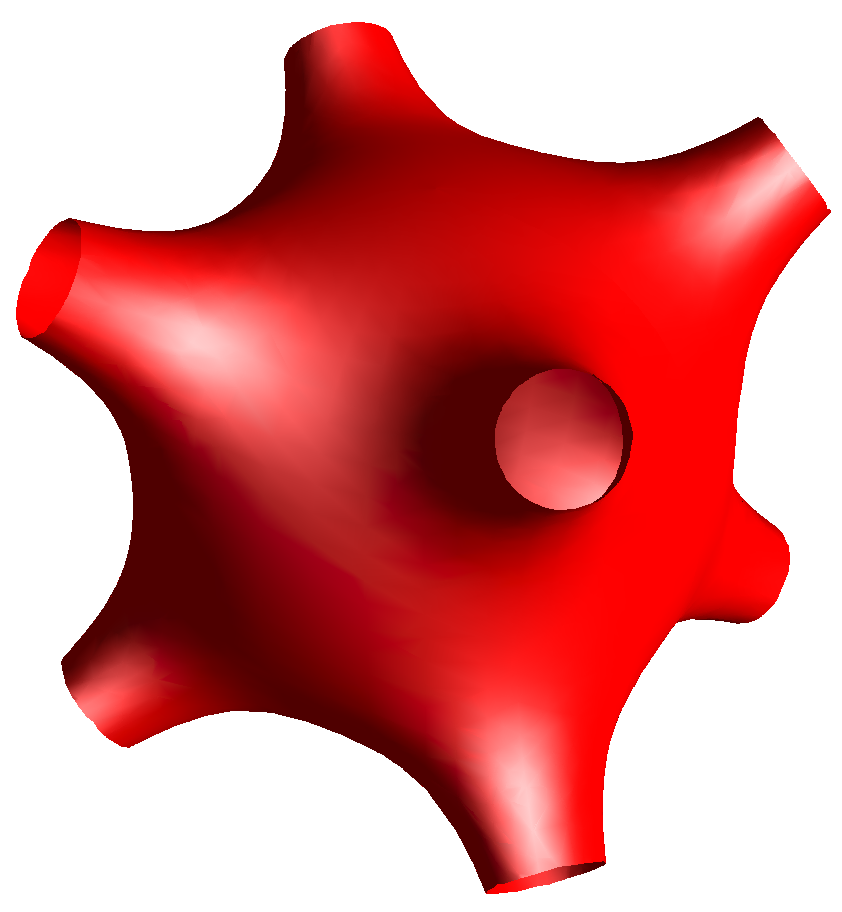}}\qquad\qquad
  \subfloat[A slice through the 24-cell solution]{\label{24cell}
    \includegraphics[height=2.5in]{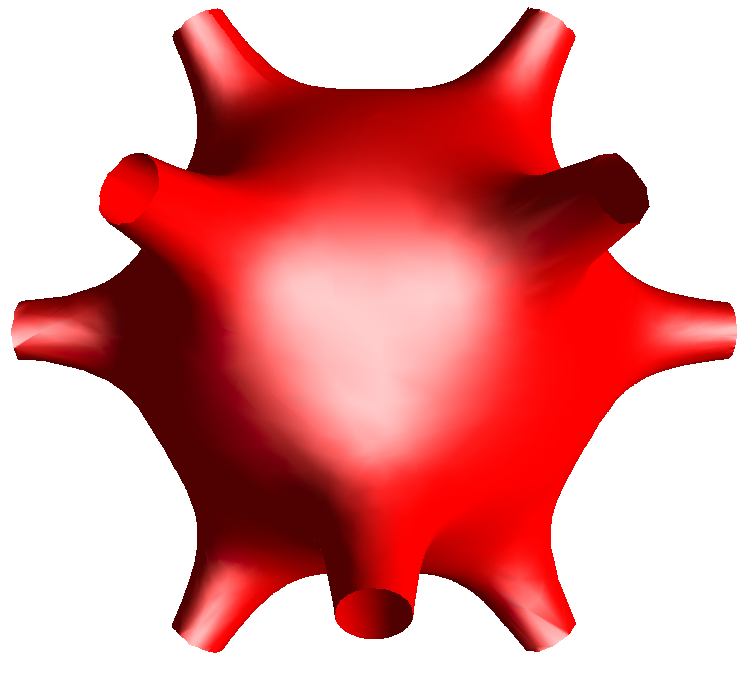}}\\
  \subfloat[A slice through the 120-cell solution]{\label{120cell}
    \includegraphics[height=2.5in]{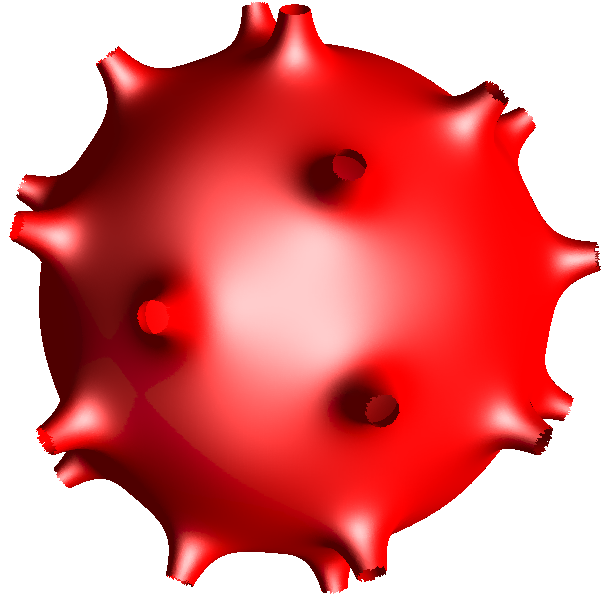}}\qquad\qquad
\subfloat[A slice through the 600-cell solution]{\label{600cell}
    \includegraphics[height=2.5in]{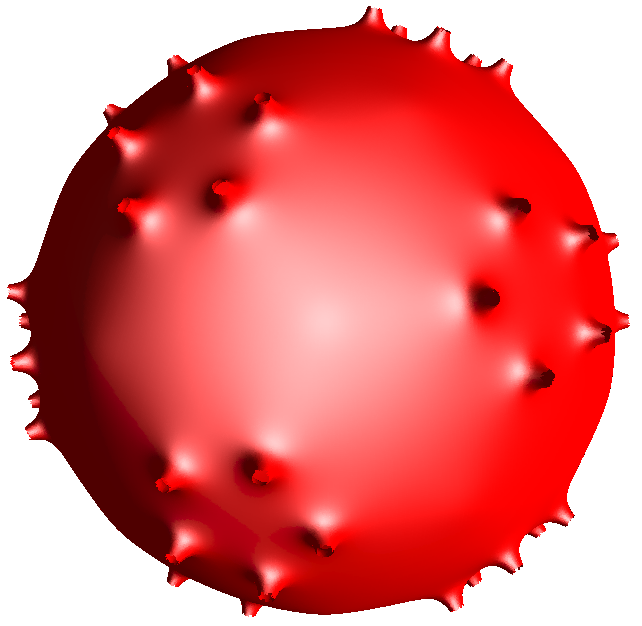}}
  \caption{Slices through the hypersurfaces of the discrete lattice
    solutions.  The distance from the centre is proportional to $\psi$.
    The tubes correspond to the locations of the masses.  These
    objects become more and more spherical as the number of sources in
the lattice increases.}
\end{figure*}

We can construct the other discrete models, made using 16, 24, 120
and 600 equally spaced masses, by proceeding in a similar way to 
the cases discussed in detail above.  We shall not present the details
of these constructions here, but to help with their visualisation we 
display slices through these structures in Figures \ref{16cell}-\ref{600cell}.  
It can be seen that as the number of masses increases,
the shape of each of these structures becomes increasingly spherical, while
the tubes become thinner.  This corresponds to the spacetime
approaching homogeneity as the number of masses is increased.
We will use the results obtained  from studying these larger lattice
models in the sections that follow.

\section{Location of the Horizons}
\label{Sec4}

To qualify as cosmological solutions we require that the discrete 
models considered here avoid having any overlap in the horizons corresponding to
different masses. We therefore need to investigate the positions 
of the horizons in the models discussed in the previous section, in
order to see if this criterion is met. In these 
models the location of the event horizon can be approximated by
marginally trapped surfaces \cite{Gibbons} (such surfaces give the
exact locations of the event horizons if the spacetime is static).

To find these trapped surfaces let us consider a surface given by some
function $\chi=\chi(\theta,\phi)$. This surface has geometry
\begin{eqnarray*}
d \sigma^2 
&=& \psi^4 (\chi^2_{,\theta}+\sin^2\!\chi ) d\theta^2
+2 \chi_{,\theta}
\chi_{,\phi} \psi^4 d \theta d \phi
 \\&&+\psi^4
(\chi^2_{,\phi}+\sin^2\!\chi \sin^2\!\theta ) d \phi^2,
\end{eqnarray*}
and the unit normal to it is
\begin{equation*}
n_{\mu} =  \frac{\psi^2 \sin \chi \sin \theta}{\sqrt{\sin^2\!\chi
    \sin^2\!\theta+\sin^2\!\theta \chi^2_{,\theta} +\chi^2_{,\phi}}}
\left( 1, -\chi_{,\theta}, -\chi^2_{,\phi} \right),
\end{equation*}
where commas in subscripts denote partial differentiation.
The extrinsic curvature of this surface is then
$K_{\mu \nu} = n_{\mu ; \nu}$, where the covariant derivative here is with
respect to the metric of the 3-space.  Transforming to
coordinates $a,b,$... on the 2-space, using
$K_{ab}=\frac{\partial x^{\mu}}{\partial x^a} \frac{\partial
  x^{\nu}}{\partial x^b} K_{\mu \nu}$,
the trapped surfaces are those that have $K=\gamma^{ab}K_{ab}=0$,
where $\gamma^{ab}$ is the contravariant induced metric on the 2-space.  This
condition is satisfied if
\begin{eqnarray}
&&4 \sin^2\!\chi \sin \theta (1+\chi_{\vert \phi}^2)(1+\chi_{\vert \theta}^2) \left( \chi_{\vert \phi}
  \psi_{\vert \phi} + \chi_{\vert \theta} \psi_{\vert \theta}- \psi_{, \chi}\right)
 \nonumber \\ \nonumber &&+\psi \sin \theta \left[ \left(1+\chi_{\vert
     \phi}^2 +\chi_{\vert \theta}^2
  \right) \chi_{\vert \phi \phi} + \left( 1+\chi_{\vert \phi}^2
  \right) \chi_{\vert \theta \theta} \right]
\nonumber \\ \nonumber &&
+\psi \cos \chi \sin \theta \big[ (1+\chi_{\vert \phi}^2) \chi_{\vert
    \theta}^2 +\chi_{\vert \phi}^2 (1+\chi_{\vert \phi}^2+\chi_{\vert
    \theta}^2) \sin \theta \nonumber \\ \nonumber &&\qquad \qquad
   \qquad \qquad \;\; -\sin \chi (2+3 \chi_{\vert \phi}^2 +(3+4
  \chi_{\vert \phi}^2) \chi_{\vert \theta}^2) \big]\\
&&= 
-\psi \sin \chi \cos \theta \chi_{\vert \theta} (1+\chi_{\vert \theta}^2),
\label{long}
\end{eqnarray}
where for compactness we have introduced the notation
$\chi_{\vert \theta}\equiv\frac{1}{\sin\chi} \chi_{,\theta}$, and
$\chi_{\vert \phi}\equiv\frac{1}{\sin\chi \sin \theta} \chi_{,\phi}$.
The positions of the horizons are then approximated by the solution to
this equation.

In practise Eq.~(\ref{long}) is not easy to solve, but we can obtain
approximate solutions by looking for minimal surfaces of constant
$\chi$, for which Eq.~(\ref{long}) reduces to
$(\psi^4\sin^2\!\chi)_{,\chi}=0$.  We expect this to be a good
approximation for all the models considered here, and for its 
accuracy to increase as the number of masses in the lattice is
increased.  To establish this result we calculate the area of the
horizon of each mass using our approximation, $A_\text{min}$, as well
as the horizon area of a Schwarzschild black hole with an equal proper
mass, $A_\text{S}$.  We then calculate the ratio
$(A_\text{min}-A_\text{S})/A_\text{S}$, which we have displayed in
Table \ref{table7}.  As expected, the difference is small, and
decreases as the number of masses is increased.

To check that the horizons are not overlapping we compare $\chi_{min}$ with
half the separation between neighbouring sources, ${\Delta\chi}$.
This is also displayed for each of our 6 discrete models in Table
\ref{table7}.  It can be seen that $\chi_{min}$ is always less than
half ${\Delta\chi}$, and that it decreases as the number of
masses is increased.  Together with the small values of
$(A_\text{min}-A_\text{S})/A_\text{S}$ this provides a good indication that
the horizons of the masses in our models do not intersect at the
maximum of expansion, thus ensuring that our discrete models satisfy a
necessary condition to qualify as cosmological models\footnote{We
  leave aside for now the more complicated question of whether there are
  additional horizons that could encompass two or more masses.  We
  note only that this would appear to be unlikely, given that in every
  case the masses are separated by multiple horizon distances.}.

\begin{table}[htb]
\begin{center}
\begin{tabular}{|c|l|l|}
\hline
$\begin{array}{c} \textbf{Number of}\\
  \textbf{Masses} \end{array}$  &
\large{\bf{$\frac{A_\text{min}-A_\text{S}}{A_\text{S}}$}} &
\large{\bf{$\frac{\chi_\text{min}}{\Delta \chi/2}$}}\\
\hline
\vspace{-6pt} &  &  \\
$5$ & $\hspace{2pt}1.55\times 10^{-6}$  & $\hspace{2pt}0.428$ \\
$8$ & $\hspace{2pt}7.15\times 10^{-9}$  &  $\hspace{2pt}0.268$ \\
$16$ & $\hspace{2pt}3.46\times 10^{-9}$  &  $\hspace{2pt}0.173$ \\
$24$ & $\hspace{2pt}9.95\times 10^{-13}$  &  $\hspace{2pt}0.110$ \\
$120$ & $\hspace{20pt}$--  &  $\hspace{2pt}0.0330$ \\
$600$ & $\hspace{20pt}$--  &  $\hspace{2pt}0.0147$ \\
\hline
\end{tabular}
\end{center}
\caption{The fractional difference between our estimate of the horizon
size, $A_\text{min}$, and the horizon size of a Schwarzschild black
hole with the same proper mass, $A_\text{S}$.  Also displayed is our
estimate of the fraction of the distance to the point half-way
between masses that the horizon reaches, for each of our discrete models. Dashes indicate numbers smaller than our numerical precision.}
\label{table7}
\end{table}

\section{Backreaction in dust models}
\label{Sec2}

We shall now employ our exact discrete solutions in order to study
backreaction in dust-filled cosmological models. This will be done by comparing the scale of discrete and continuous models on their time-symmetric hypersurfaces, at the maximum of expansion.

Here we will make use of the notion of `proper mass' in order to
compare our lattice models with the dust-filled ($k=+1$) FLRW solutions.
That is, for a given lattice we will calculate the total proper mass
of all the sources in the lattice, and we will then compare this
lattice to an FLRW solution with the same total proper mass.  
{\bfx The motivation for this procedure is that in both cases, the
  proper mass corresponds to locally measured masses. {\bfx
    Specifically, for the FLRW models, the proper mass for a given
    region can be defined as the integral of the energy density over
    that region. 
For the lattice models,
the mass within each cell is identified with the proper mass of the
source it contains, so that the total proper mass of the lattice is given by the sum of the proper masses of all the sources within it.}}

The geometry of the time-symmetric hypersurface of maximum expansion
in ($k=+1$) FLRW solutions is now given by Eq.~(\ref{FLRWmax}).  On the other hand, the metric
corresponding to the 3-sphere containing $n$ discrete masses of equal
size is given by Eq.~(\ref{sphere_general}), with $\tilde m_{i+1}= \tilde m_i$ for
every $i<n$.  The value of $\tilde m_i$ can then be related to the proper mass
of each of the objects using the procedure outlined in Section
\ref{msm}.  In order to compare continuous and discrete solutions we
then only need to make sure that the mass of the continuous solutions (defined as
the constant $M= \rho V$ where $V$ is the total spatial volume of the
universe) is the same as the sum of the proper masses in the discrete solutions.

In the discrete solutions, consisting of $n$ objects each with
proper mass $m$, the total mass is clearly just $M=nm$.  In the
FLRW solutions the volume of a spatial section of constant $t$ is given
by $V = {2 \pi^2 a^3}$, where $a=a(t)$ is the scale factor. Thus
the energy density for a ($k=+1$) FLRW solution with the same total mass 
($M$) is given at its maximum of expansion by 
\be\label{rhozero}
\rho(t_0) = \frac{M}{V} =\frac{M }{ 2 \pi^2 a^3(t_0)}.
\ee
Recalling that in a ($k=+1$) FLRW solution the maximum of expansion
occurs when $a^2 = 3/(8\pi \rho)$, the line-element
(\ref{FLRWmax}) can be written as
\begin{equation}
\label{FLRW8b}
dl^2 = \frac{16 M^2}{9 \pi^2} \left( d\chi^2+\sin^2\!\chi \,
d\theta^2 +\sin^2\!\chi \sin^2\!\theta \, d\phi^2 \right).
\end{equation}
We can now compare this geometry with the corresponding
discrete geometry given by Eq.~(\ref{sphere_general}).

To proceed we require a measure of the global scale for both the
discrete and the continuous solutions. In the continuous case it is clear 
what this measure should be, as there is only one scale in the geometry
(the curvature of the 3-sphere). For the discrete solutions, on the other
hand, the length of a curve of fixed angle on the
3-sphere will depend on its particular position,
as the geometry of the space is inhomogeneous in these solutions.  We
must therefore proceed with some care. Here, for the discrete solutions, 
we propose two possible measures of the `size' of the space.
These are:
\begin{enumerate}
\item[D1] The line-element, $dl$, of a curve at a vertex of the lattice.
  These are clearly distinguished positions within the lattice,
  corresponding to the points that are furthest from all masses.
\item[D2] The length of the edge of a cell.  Again, this is clearly a
  preferred curve within the lattice.
\end{enumerate}

\noindent
Both of these proposed definitions of size are aimed at trying to
identify the scale of the lattice structure that the masses occupy, as
this is the closest thing to a ``background'' that
exists in these solutions.  Identifying the size of the lattice was also
what was attempted by Lindquist and Wheeler in their
approximate solution with the same configuration of masses \cite{LW}.  Within this
context, Definition D1 will turn out to be the most conservative
possible comparison of the scales of the discrete and continuous
solutions, and Definition D2 will be found to be not very different.

\begin{figure}
\hspace{-10pt}
\includegraphics[height=3.5in]{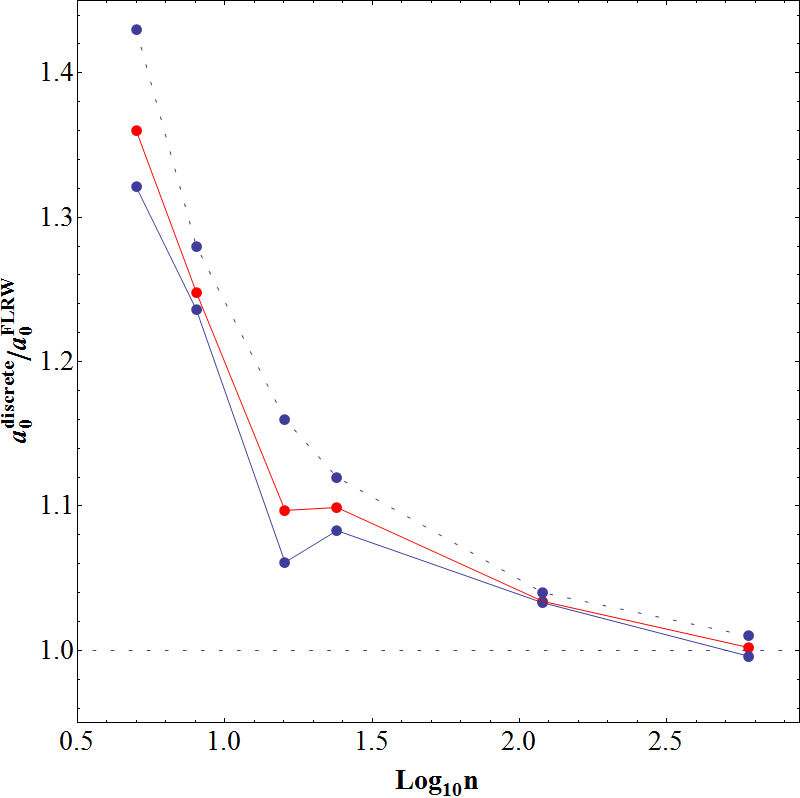}
\caption{
{\bfx The scale of the discrete solutions as a fraction of the
scale of the continuous solutions, when the total mass in the
continuous model is taken to be equal to the total proper mass of the
discrete model.} The full blue curve shows the result of using
Definition D1 for the scale of the discrete solution, and the red
curve shows the result for Definition D2. The dotted blue curve which
represents the approximate Lindquist-Wheeler solution is shown for comparison.
}
\centering{}\label{LWfig}
\end{figure}

Let us first consider Definition D1.  In this case it can be seen that
the ratio of the line-elements of curves that cover the same angle on
the 3-sphere in the continuous and the discrete solutions
is given by
\be
\label{rat1}
\frac{dl^{\rm discrete}}{dl^{\rm FLRW}} = \frac{3
  \pi \tilde{m}_i}{16 n m} \left( \sum_{i=1}^n f_i^{-1}\right)^2.
\ee
Due to the fact that $\tilde{m}_i$
can be shown to be in direct proportion to
$m$, it can be seen that this expression is, in fact, independent of
$m$.  It is therefore specified uniquely by
the structure of the lattice (i.e.\ by $n$, and the functions $f_i$ corresponding to the lattice
in question).  

The ratio of scales in Eq.~(\ref{rat1}) is a function of $\chi$,
$\theta$ and $\phi$, and so varies depending on which point in the
discrete solution we wish to consider. Using Definitions D1 and D2,
we find the results displayed in the third and fourth columns of 
Table \ref{mmtable}, respectively. Here $a_0^{\rm discrete}$ refers to the scale of discrete solution, as defined using D1 or D2.
These results are also displayed graphically in Fig.~\ref{LWfig}, for each of
our lattices.  It can be seen that the scales corresponding to
the discrete solutions is always larger than that of the corresponding
continuous solution, with the only exception being the scale of the
solution with $600$ masses, when Definition D1 is used.

\begin{table}[b]
\begin{center}
\begin{tabular}{|c|c|c|c|}
\hline
$\begin{array}{c} \textbf{Cell}\\
\textbf{Shape} \end{array}$
& $\begin{array}{c} \textbf{Number of}\\
  \textbf{Masses} \end{array}$  & 
$\bf{ \left( \frac{a_0^{\rm discrete}}{a_0^{\rm FLRW} } \right)_{D1}} $
& $\bf{ \left( \frac{a_0^{\rm discrete}}{a_0^{\rm FLRW} } \right)_{D2}} $\\
\hline
 Tetrahedron & 5 & 1.321 & 1.360 \\
 Cube & 8 & 1.236 & 1.248 \\
 Tetrahedron & 16 & 1.061 & 1.097 \\
 Octahedron & 24 & 1.083 & 1.099 \\
 Dodecahedron & 120  & 1.033 & 1.034 \\
 Tetrahedron & 600 & 0.996 & 1.002 \\ 
\hline
\end{tabular}
\end{center}
\caption{{\protect{\textit{
{\bfx The scale of the discrete solutions as a fraction of the
scale of the continuous solutions, when the total mass in the
continuous model is taken to be equal to the total proper mass of the
discrete model.} The two definitions of scale
in the discrete solutions have both been calculated.
}}}}
\label{mmtable}
\end{table}

Figure \ref{LWfig} shows that $a_0^{{\rm discrete}}/a_0^{{\rm
    FLRW}}$ approaches $1$ as $n$ becomes large, but that this
approach is not exactly monotonic.
In particular, the ratio for the lattice with 16 masses
is a little lower than might have been the case for a smooth curve.
This behaviour is likely to be a consequence of the fact that the
shape of the cells used in different tessellations are different.
For the lattice solutions made from tetrahedra (i.e.\ those with $n=5,16,600$ masses), 
for example, the approach to the FLRW limit does appear to be
monotonic.  For the sake of comparison, we have also shown in
Fig.~\ref{LWfig} the  ratios obtained using the approximate solutions
of Lindquist and Wheeler \cite{LW}. It can be seen that the 
ratios given by the approximate solutions compare well with those of the
exact solutions.

The outcome of this comparison demonstrates that, for large numbers of
regularly arranged masses at their maximum of expansion,
this measure of backreaction in dust-filled models is small.

\section{Effective Masses and Interaction Energy}
\label{Sec3}

\begin{figure}
\includegraphics[width=3.3in]{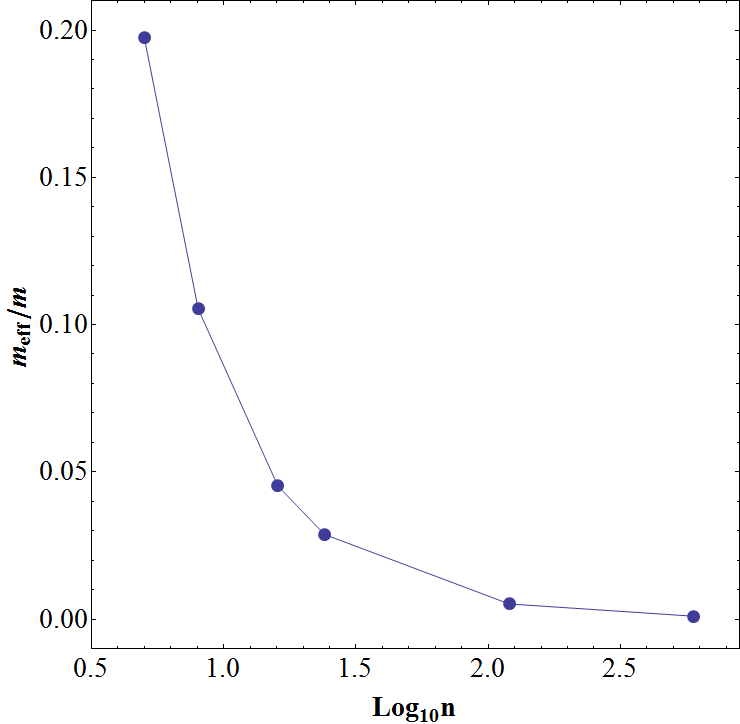}
\caption{The ratio of the effective mass, $m_{\rm eff}$, to the proper
  mass, $m$, for each lattice solution.}
\centering{}\label{LWfig2}
\end{figure}

{\bfx An advantage of the discrete models we consider is that they provide us with a
  framework within which it is possible to discuss the role of
  inter-particle interactions in cosmology.  Given that these are the
  interactions that govern the structure and evolution of the real
  universe}, we consider it to be an important undertaking to obtain a
  deeper understanding of them.  In particular,
{\bfx interactions are ignored when
approximating the universe as being filled with dust, and so {\bfx these
considerations} potentially allows us further insights into the behaviour
of spacetimes filled with discrete objects.}

{\bfx Here we will begin by comparing the notions of {\it proper mass} and {\it
 effective mass} for each of our discrete solutions, as defined in Section 
\ref{msm}.  We remind the reader that our definition of effective mass is 
based on analogy with the asymptotically flat case studied by Brill and 
Lindquist \cite{Brill-Lindquist-63}, as it is problematic to define it  
operationally  in the closed spherical settings we are currently studying. 
Now,} the values of these {\bfx masses} can be set to any given value for each lattice,
but the ratio of effective mass to proper mass must take a
particular constant value for any given configuration.  We display
this ratio for each of the six possible tessellations in
Table \ref{table-ratio}, and show it graphically in Fig.~\ref{LWfig2}. 
It is clear that the ratio of effective mass to proper mass increases
as the number of masses in the lattice is decreased.  This is
due to decreasing contributions from the interaction energies
between particles, which correspondingly increase the value of the
effective mass of each source. 

\begin{table}[t]
\begin{center}
\begin{tabular}{|c|c|c|}
\hline
$\begin{array}{c} \textbf{Cell}\\
\textbf{Shape} \end{array}$
& $\begin{array}{c} \textbf{Number of}\\
  \textbf{Masses} \end{array}$  &
$\bf{\frac{m_{\rm eff}}{m}} $\\
\hline
 Tetrahedron & 5 &  0.20\\
 Cube & 8 &  0.11  \\
 Tetrahedron & 16 &  0.045  \\
 Octahedron & 24 &  0.029 \\
 Dodecahedron & 120  &  0.0052 \\
 Tetrahedron & 600 &  0.0010 \\
\hline
\end{tabular}
\end{center}
\caption{\textit{The ratio of the effective mass, $m_{\rm eff}$, to
    the proper mass, $m$, for each lattice.}}
\label{table-ratio}
\end{table}

\begin{figure}
\includegraphics[width=3.3in]{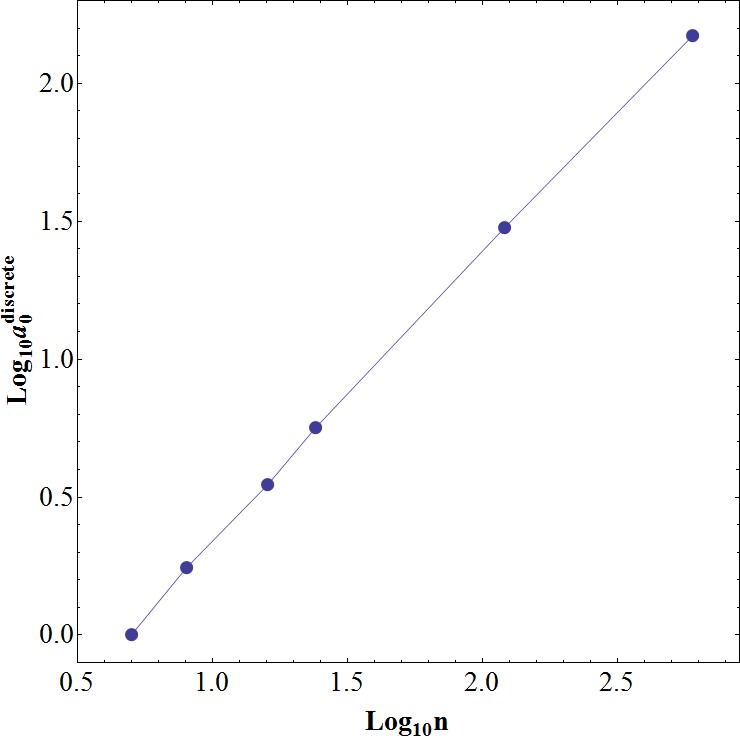}
\caption{The scale of lattice solutions with the same total effective
  mass, normalised so that $a_0=1$ for the 5-cell, and using D1 for the
scale of the discrete solutions.}
\centering{}\label{LWfig2a}
\end{figure}

In Fig.~\ref{LWfig2a} we compare the scale of the hypersurface of
maximum expansion for each of our discrete solutions, when the total
effective mass of the lattice is kept constant.  As the effective mass
includes the inter-particle interaction energies, this procedure of
comparing the scale of solutions with the same total effective mass is
equivalent to enforcing the condition that the lattices being compared
should contain the same total energy (that is, the same total proper mass plus
interaction energies).  Fig.~\ref{LWfig2a} then shows that the scale of
the hypersurface of maximum expansion grows approximately linearly with
the number of masses, $n$.  Changing the method of determining the
scale of the discrete solutions from D1 to D2 has very little effect.

The results in Fig.~\ref{LWfig2a} can be understood by considering Fig.~\ref{LWfig2} -- as the
number of masses decreases the ratio of the magnitude of interaction energies to proper
mass also decreases.  The interaction energies, however, are negative,
so if we consider a thought experiment in which we deform the 8-cell into the 5-cell
we have to {\it increase} the total amount of energy available for
interactions.  The only source of energy in the solution is the proper
mass of the particles, and so a lattice with fewer masses must have a
smaller total proper mass if the total energy in the system is to
remain unchanged.  Smaller total proper mass corresponds to a smaller
scale for the hypersurface of maximum expansion, {\bfx as was shown in
Section \ref{Sec2}}.  This result
suggests that one can substantially increase the scale of a
closed space by dividing up the mass in that space into smaller
packets, although we have strictly only shown this is the case on
time-symmetric hypersurfaces\footnote{For a dynamical cosmological
model we no longer necessarily have energy conservation, and so the
situation could be more complicated.}
{\bfx It should also be noted that this result relies on the definition of
  effective mass that we have generalized from the asymptotically flat
  setting in order to define the interaction energies.  The validity
  of this approach is difficult to confirm operationally in a
  compact space, and so the interpretation of the 
dramatic increases in scale seen here should be treated with care.}  

{\bfx Such increases in scale} are entirely absent in dust models, as interaction energies
are not included in the energy budget in that case.  This does
not, however, invalidate any of the conclusions of Section
\ref{Sec2}.  It remains the case that back-reaction is small in models
where interaction energies are ignored.  What has changed for the
lattices in Fig.~\ref{LWfig2a} is that the proper mass in each lattice
is increasing as the number of masses is increased, so the
dust-dominated FLRW solution that one should compare with these models is also changed.

\section{Conclusions}
\label{conclusion}

We have studied the emergence of spatially closed
Friedmann solutions from inhomogeneous solutions that contain increasing
numbers of regularly arranged discrete masses in topological
3-spheres.  This has been done using exact methods, 
and therefore allows for exact measures of backreaction in dust
dominated cosmological models.

More specifically, by considering the instantaneously static
hypersurfaces at the moment of maximum expansion we have found that
universes that contain only a small number of mass concentrations
($\lesssim 10$) can be $10\%$ or more larger than the
corresponding dust dominated FLRW solutions of Einstein's equations.
However, for universes that contain very many masses ($\gtrsim 100$)
there is very little quantitative difference ($\lesssim 1\%$) between
the scale of maximum of expansion in the discrete and continuous
solutions.  These results are based on comparing dust dominated continuous
models (which ignore the inter-particle interaction energies) with
momentarily static discrete models that contain the same total `proper mass' (which also
ignores the interaction energies between the different masses). 
{\bfx While backreaction in this case is quite small, it may well be that the inter-particle
  interactions will come into play in a more substantive way when we
  allow for dynamics.  That is, backreaction could still
  have a big effect when we consider the evolution of the universe. 
}

The results we find are in good keeping with the approximate solutions
of Lindquist and Wheeler that use a gravitational analogue
of the Wigner-Seitz construction \cite{LW}, as is demonstrated
graphically in Fig.~\ref{LWfig}.  {\bfx They also present us with exact
  expressions for the change in scale of cosmological solutions that
  has been predicted using Zalaletdinov's averaging scheme
  \cite{Coley,vdH,Zala}.}
More generally, we believe our study could allow some insight into
the problem of how averaging should be performed in relativistic
cosmology. To date, most studies on this subject have used either
highly symmetric exact solutions that contain a perfect fluid, or have
considered small fluctuations around an FLRW ``background'' geometry.
The former approach is strongly limited by the high degree of symmetry required in the
solutions, while the latter is limited to geometries that are already
necessarily close to FLRW.  The models we have constructed here suffer
from neither of these shortcomings, as they admit no Killing vectors, and
do not require the assumption of an FLRW geometry, either as a boundary
condition for the inhomogeneities, or as a background geometry.  As
such it offers a new laboratory for testing ideas about inhomogeneity,
averaging, and backreaction in cosmology.

With the notion of effective mass employed here,
we find the interaction energy grows rapidly as the number of masses
in the lattice, and comes to dominate in the limit of very many
masses.  This means that if we compare lattices with different
numbers of masses, but with the same total energy (including
interaction energy), then the scale of the hypersurface of maximum
expansion increases dramatically with increasing number of masses.
{\bfx As was discussed above, however, the interpretation of this effect
requires some care, as it is based on a generalization of the definition of effective mass given
by Brill and Lindquist in asymptotically flat settings. Nevertheless,
this effect} is entirely neglected when treating the matter content of
the universe as dust.

Finally, in this paper, we have confined ourselves to the comparison of the discrete and
continuous models on time symmetric hypersurfaces. Clearly the next step would be to make
a detailed comparison of the full evolution of these models. We shall
return to this question in future publications.  We also note that,
although we have only considered regular arrangements of masses in
this paper, the formalism we have used allows for the possibility of
considering much more complicated distributions.
\newline
\;
\newline
{\bf Note added}. After submission of our manuscript the following related work appeared: \cite{related1,related2,related3}. The first of the these papers performs a numerical analysis of a spatially flat lattice of black holes \cite{related1}, while the second performs a perturbative analysis of a similar situation \cite{related2}. The third study finds exact initial data for a lattice of eight black holes in a space with spherical topology (using the same method presented in this paper), and then proceeds to numerically evolve this data \cite{related3}.
\newline
\;
\newline
{\bf Acknowledgements}.
\newline

TC acknowledges the support of the STFC and the BIPAC. 
TC and RT thank the hospitality of UCT for a visit in February 2011, and TC and KR thank the AU, QMUL for a 
visit in October 2011. We are grateful to George Ellis, Helena Engstr\"om, Pedro
G. Ferreira, Daniele Gregoris, Roy Maartens and Bruno Mota for helpful comments and discussions. 



\end{document}